\documentclass[reprint,superscriptaddress,nofootinbib,twocolumn,amsmath,amssymb,aps,pra]{revtex4-2}
\usepackage{graphicx}
\usepackage{braket}
\usepackage[colorlinks, citecolor=red, urlcolor=red, bookmarks=false, hypertexnames=true]{hyperref}
\usepackage{dcolumn}
\usepackage{bm}
\usepackage{hyperref}
\usepackage{mathrsfs}

\begin{document}
\preprint{APS/123-QED}
\title{
High angular momentum coupling for enhanced Rydberg-atom sensing in the VHF band}
\author{Nikunjkumar Prajapati}\affiliation{National Institute of Standards and Technology, Boulder, CO 80305, USA}

\author{Jakob W. Kunzler}\affiliation{Naval Information Warfare Center Atlantic, Hanahan, SC, 29410}

\author{Alexandra B. Artusio-Glimpse}\affiliation{National Institute of Standards and Technology, Boulder, CO 80305, USA}

\author{Andrew Rotunno}\affiliation{National Institute of Standards and Technology, Boulder, CO 80305, USA}

\author{Samuel Berweger}\affiliation{National Institute of Standards and Technology, Boulder, CO 80305, USA}

\author{Matthew T. Simons}\affiliation{National Institute of Standards and Technology, Boulder, CO 80305, USA}

\author{Christopher L. Holloway}\affiliation{National Institute of Standards and Technology, Boulder, CO 80305, USA}

\author{Chad M. Gardner}\affiliation{Naval Information Warfare Center Atlantic, Hanahan, SC, 29410}

\author{Michael S. Mcbeth}\affiliation{Naval Information Warfare Center Atlantic, Hanahan, SC, 29410}

\author{Robert A. Younts}\affiliation{Naval Information Warfare Center Atlantic, Hanahan, SC, 29410}

\date{\today}

\begin{abstract}
Recent advances in Rydberg atom electrometry detail promising applications in radio frequency (RF) communications. Presently, most applications use carrier frequencies greater than 1~GHz where resonant Autler-Townes splitting provides the highest sensitivity. This letter documents a series of experiments with Rydberg atomic sensors to collect and process waveforms from the automated identification system (AIS) used in maritime navigation in the Very High Frequency (VHF) band. Detection in this band is difficult with conventional resonant Autler-Townes based Rydberg sensing and requires a new approach. We show the results from a new method called High Angular Momentum Matching Excited Raman (HAMMER), which enhances low frequency detection and exhibits superior sensitivity compared to the traditional AC Stark effect. From measurements of electromagnetically induced transparency (EIT) in rubidium and cesium vapor cells, we show the relationship between incident electric field strength and observed signal-to-noise ratio and find that the sensitivity of the HAMMER scheme in rubidium achieved an equivalent single VHF tone sensitivity of $\mathrm{100~\mu V/m/\sqrt{Hz}}$.  With these results, we estimate the usable range of the atomic vapor cell antenna for AIS waveforms given current technology and detection techniques.
\end{abstract}

\maketitle

\section{Introduction}
Rydberg atom based electric field sensors have grown in popularity and utility over the past decade. With their ability to offer field measurements traceable to the international system of standards~\cite{holl1,si_trace,SI_standard} and beat the Chu limit~\cite{PhysRevLett.121.110502}, their application in low SWaP (size, weight, and power) and fixed environments has drawn interest from the antenna and physics community ~\cite{9748947}.  From their first demonstration for traceable electric field measurements~\cite{first_ryd_elet_shaff,holl1}, many new applications have risen \cite{9748947}. This includes and is not limited to phase resolved measurements for Angle-of-Arrival (AOA) ~\cite{aoa}, quadrature amplitude modulation (QAM) reception~\cite{8778739,doi:10.1063/1.5088821}, wide scan range spectrum analyzers ~\cite{PhysRevApplied.15.014053}, radio frequency (RF) power standards ~\cite{pow_stand}, alternating and direct current (AC/DC) voltage standards~\cite{doi:10.1116/5.0097746}, video reception~\cite{doi:10.1116/5.0098057}, and many more applications~\cite{9748947, doi:10.1063/1.5099036,Carr:12,7812705,rydberg_array,Prajapati:18,terehertz_imaging}.

Rydberg atoms boast a large polarizability due to the separation of the nucleus and the highly excited electron~\cite{gallagher_book}. This makes them highly susceptible to external electric fields. Furthermore, one can tune the RF resonance of the Rydberg atoms by selecting the principle quantum number $n$ of the Rydberg state of the atom's excitation~\cite{7812705}. However, for low-frequency applications, this method requires the tuning to very high $n$ where many deleterious broadening mechanisms occur, such as atomic collisions, charge effects, and various decoherence mechanisms~\cite{Fan_2015}. For broadband applications of the Rydberg atom sensor, an off-resonant AC Stark effect is more suitable~\cite{waveguide_SA}. Furthermore, this allows for detection of frequencies in the Very High Frequency (VHF)  (50 MHz to 300 MHz) and ultra-high frequnecy (UHF) (300 MHz to 900 MHz) bands and below without suffering the consequences of high $n$ tuning. Yet, with the off-resonant Stark shift, there are limitations to the sensitivity of the Rydberg atoms. Groups have demonstrated weak field, satellite signal detection with the use of a high gain antenna, low noise amplifiers, a wave-guide coupled vapor cell, and a dressed ladder system to maintain a low $n$ all to measure XM band (lower) frequency signals~\cite{satellite_XM}. But to show true supremacy as an antenna not bound by the Chu limit, other methods are needed.

\begin{figure}[t]
    \centering
    \includegraphics[width = .4\textwidth]{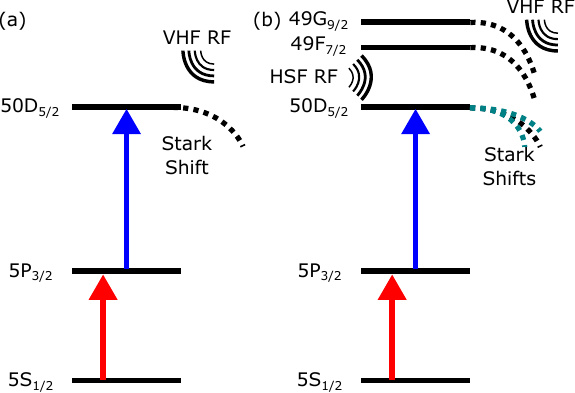}
    \caption{(a) Level diagram showing the interaction of AC Stark shifting measurements. (b) Level diagram showing the interaction of coupling in the high angular momentum F and G states that cause mixing and enhancement of the measurement.}
    \label{fig:level_diagram}
\end{figure}

We demonstrate the use of ``Rydberg state engineering", similar to what is mentioned in ~\cite{PhysRevApplied.19.044049}. In this paper, we demonstrate VHF detection in two ways for comparison. In the first case, we utilize Stark shifting of resonances as represented in Fig.~\ref{fig:level_diagram} (a). In the second case, we use a method that we developed, High Angular Momentum Matched for Exciting Raman (HAMMER) method, shown in Fig.~\ref{fig:level_diagram} (b). This method involves the use of a dressing RF field that couples the Rydberg state to a nearby higher angular momentum Rydberg state through a resonant super high frequency (SHF) RF transition. Then, by applying a strong VHF local oscillator (LO), we Stark shift the F and G Rydberg states to bring the F to G transition into resonance with the VHF signal field (discussed more in Section~\ref{sec:hammer}). The boost in sensitivity comes from the higher angular momentum coupling in from the now resonant dressing. For example, in rubidium (Rb) the polarizability of the 50D$_{5/2},m_j=5/2$ state is 0.02~MHz(V/m)$^{-2}$ while the polarizability of the 49F$_{7/2},m_j=1/2$ is 1.5~MHz(V/m)$^{-2}$ and the 49G$_{9/2},m_j=1/2$ is 5.5~MHz(V/m)$^{-2}$. This difference is less pronounced in the cesium (Cs) atoms, but still provides a benefit.

We compare the sensitivity and usable signal of the bare-state Stark shifting method to the dressed-state HAMMER method. To study the usability of these schemes for communication, we apply automatic identification system (AIS) signals that are used for maritime navigation.

\section{Stark Shifting vs. HAMMER}\label{sec:hammer}
In practice, most Rydberg atom electric field sensor demonstrations have utilized resonant atomic effects. Additionally, there have been recent efforts to allow for the Raman coupling between several Rydberg states through two-photon RF interactions~\cite{PhysRevApplied.19.044049}. There have also been efforts utilizing Stark shift based measurements to achieve broad tunability~\cite{waveguide_SA}. Here, we demonstrate a mixture of several of these techniques that allows for engineering an atomic level structure that enhances specific sensing capabilities of the atoms. 

To provide an example and explain the mixed state effects we see, we focus this discussion on Rb atoms and the line shifting expected from the external fields; however, similar interactions occur in the Cs atoms as well. When we try to estimate the sensitivity of a Rydberg atom electrometry setup, we typically look at the change in transmission of the probe laser with the application of some electric field. The minimum resolvable change determines the electric field sensitivity of the sensor. The larger a shift produced by a given electric field, the better the sensitivity of the method.

In the experiment, we excite to the 50D$_{5/2}$ Rydberg state. We then apply a field resonant with the 50D$_{5/2}\rightarrow$ 49F$_{7/2}$ transition at 18.56~GHz. Following this, we apply a 162~MHz VHF local oscillator (LO) field. This has two purposes. The first is to provide a field reference for phase based measurements. The second is the bedrock for the HAMMER method to work. The field is increased in amplitude until there is a strong Stark shift on the 49G$_{9/2}$ Rydberg state such that the 49F$_{7/2}\rightarrow$ 49G$_{9/2}$ becomes resonant with the 162~MHz field. This produces a mixing of the G and F states with the D state. In this way, we engineer the states to operate in their most sensitive configuration. 
\begin{table}[t]
    \centering
    \begin{tabular}{c|c|c|c|c|c}
       m$_j$value  &  4.5  &  3.5 & 2.5 & 1.5 & 0.5 \\
       \hline
       50D$_{5/2}$     &  NA	& NA & 0.0212& 0.0425& -0.042\\
       49F$_{7/2}$    & NA   & 0.6697& 1.069& 1.328& 1.455\\
       49G$_{9/2}$    &  2.553&	3.864&	4.802&	5.365&	5.552\end{tabular}
    \caption{Polarizability (MHz(V/m)$^{-2}$) of the Rydberg states calculated using the ARC Rydberg calculator.}
    \label{tab:polarizabilities}
\end{table}

\begin{figure}[t]
    \centering
    \includegraphics[width=.48\textwidth]{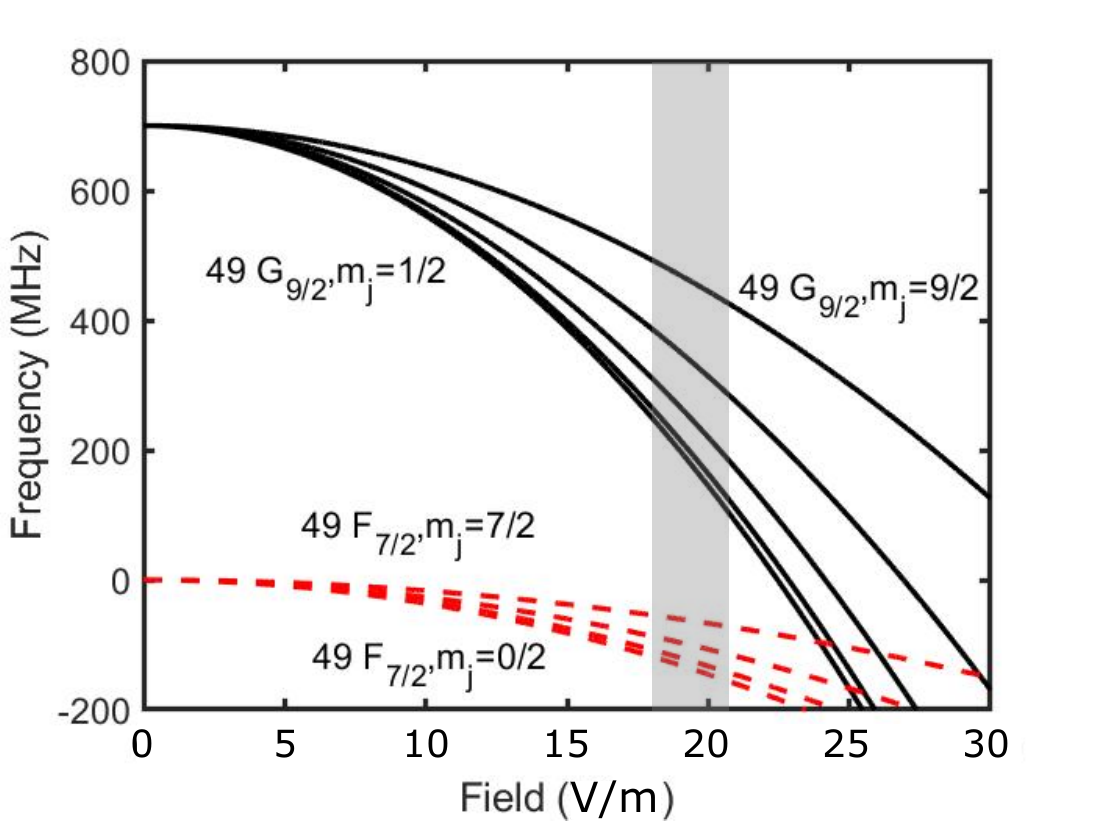}
    \caption{Stark map showing the 49F$_{7/2}$ (red dashed) and 49G$_{9/2}$ (black solid) states. The additional lines of the same color are the m$_j$ levels with lowest and highest levels labeled. The gray shaded region shows the first instance (defined by closes m$_j$ levels) where the two states are roughly 162~MHz apart.}
    \label{fig:StarkMap}
\end{figure}

The polarizabilities of the Rb Rydberg states are given in Table.~\ref{tab:polarizabilities}. These polarizabilities are calculated using the Alkali Rydberg Calculator (ARC)~\cite{vsibalic2017arc}. We plot out the AC Stark shifts for the 49F$_{7/2}$ and 49G$_{9/2}$ to show the level of field needed to achieve this method of operation in Fig.~\ref{fig:StarkMap}. As the strength of the 162 MHz LO is increased, both the 49F$_{7/2}$ and 49G$_{9/2}$ states shift, but due to the larger polarizability of the 49G$_{9/2}$ state, there is a region where the G state cross the F state. There is also a region, denoted by the gray shaded area, where the F and G state are roughly 162~MHz apart, making the interaction resonant rather than off resonant. This is the region where we operated experimentally to obtain optimal sensitivity. The field strength needed to move the two states to within 162~MHz of each other is roughly 20~V/m. We calibrate this measurement by measuring the Stark shift of the 50D$_{5/2},m_j=1/2$ Rydberg state. 

\section{Experimental Apparatus}
\begin{figure*}[!ht]
    \centering
    \includegraphics[width = \textwidth]{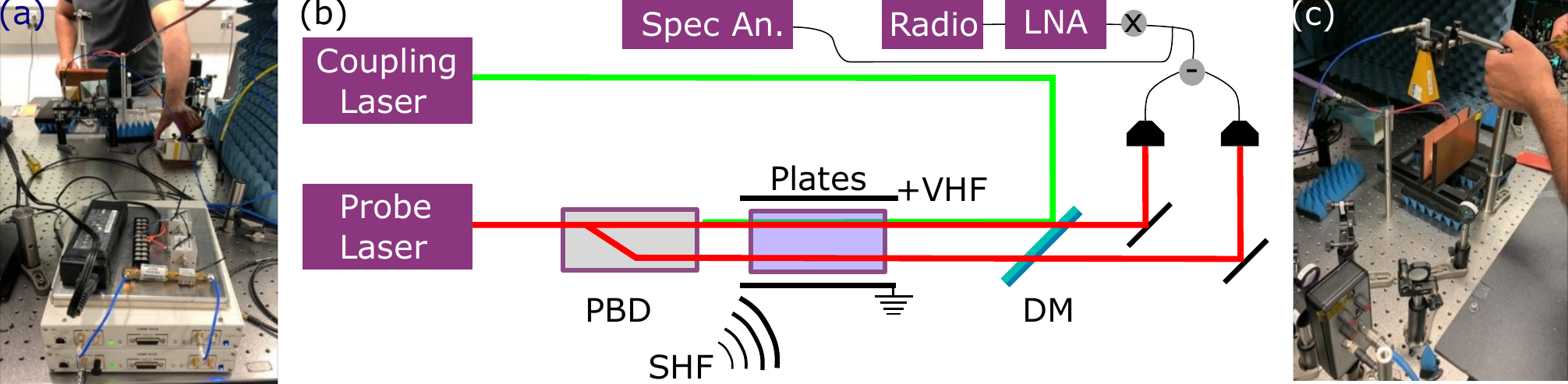}
    \caption{(a) Picture of radio, LNA, and cable connections to plates and horn. (b) Experimental schematic showing laser interactions in cell and connections. (c) Picture of setup showing plates and horn that supply fields to cell.}
    \label{fig:schematic}
\end{figure*}

\subsection{Rydberg Electric Field Detection}
We excite Rydberg atoms and probe them using two-photon EIT, as shown in Figs.~\ref{fig:level_diagram} (a) and ~\ref{fig:schematic} (b) and (c). 
For the rubidium system, we use a 780~nm probe and 480~nm coupling laser.
For the cesium system, we use an 850~nm probe and 510~nm coupling laser.
The probe laser for each setup was split using a polarizing beam displacer (PBD) to generate a reference and signal arm.  Both arms pass through the vapor cells. 
The beam sizes were approximately 1~mm on both the Cs and Rb system. 
The vapor cells used in both the Cs and Rb experiments were 25~mm in diameter, but the Cs was 25~mm long while the Rb cell was 75~mm long. 
Because our goal is not to compare Rb to Cs, but rather the effects of the high angular momentum states, we did not adjust the Rb and Cs setups to be perfectly matched.

The VHF signal field was provided by a pair of parallel copper plates measuring 10~cm by 15~cm placed in a 3D printed cradle and separated by 6~cm. 
The plates were driven by a 162~MHz AIS signal generated by the radio.
We also radiated the atoms with an LO generated by an external signal generator set to a 15~kHz frequency offset from the target AIS signal at 162.025 MHz.  The LO signal beats with the AIS in the atoms to provide a single sideband signal centered at 15~kHz at baseband.  The single sideband signal contains the modulations of the AIS signal. We additionally applied the SHF dressing field for the HAMMER measurements through a horn antenna mounted above the vapor cell, as shown in Fig.~\ref{fig:schematic} (c).

For the Cs system, the probe laser was tuned to the $6S_{1/2},F=4\rightarrow6P_{3/2},F=5$ transition and the coupling laser was tuned to the $6P_{3/2},F=5\rightarrow56D_{5/2}$ transition. The dressing field to reach the high angular momentum state is 4.07~GHz for the $56D_{5/2}\rightarrow55F_{7/2}$ transition. 
For the Rb system, the probe laser was tuned to the $5S_{1/2},F=3\rightarrow5P_{3/2},F=4$ and the coupling laser was tuned to the $5P_{3/2},F=4\rightarrow50D_{5/2}$ transition. The tuning field to reach the high angular momentum state for the HAMMER measurements is 18.5~GHz for the $50D_{5/2}\rightarrow49F_{7/2}$ transition. 

\subsection{Radio Calibration and Implementation}
For these experiments, we utilized a software-defined radio to generate multiple wave-forms, shown in Fig.~\ref{fig:schematic} (a). We calibrate the electric field generated by the radio output by determining the Stark shifting of the electromagnetically induced transparency (EIT) peak with increasing power.  By observing the peak movement for different powers of the radio, we obtain a curve to determine the received field at the atoms.  The calibration is performed by stimulating the atoms with a single tone at 162 MHz from the SDR at 100\% modulation depth. We calibrate for several analog gain levels of the SDR. By changing the gain on the SDR, we also change the Stark shift on the Rydberg state being measured with EIT. This shift is then used to calculate the incident electric field amplitude inside the test apparatus based on the tone's amplitude. Fig.~\ref{fig:calibration} in Appendix.\ref{sec:AppendCal} shows the measured calibration curves in Rb and Cs. With this calibration for a given analog gain of the SDR, we can adjust the modulation depth from 100\% to 0.001\% with a linear scaling relative to field. This
adjustment was made according to the modulation depth of the AIS signal envelope relative to the carrier wave amplitude established during the calibration process.  

AIS is a packet protocol containing National Marine Electronics Association (NMEA) navigation messages modulated with Gaussian Minimum Shift Keying (GMSK) at a symbol rate of 9600~Hz.  Class A commercial broadcasters radiate 12.5~Watts of power, and class B private crafts radiate 2~W.  AIS uses two VHF channels at 161.075 MHz and 162.025 MHz.  Most of the power spectral density is contained within the 9600~Hz bandwidth of spectrum around the carrier.  The NMEA payloads are encoded with the high-level data link control (HDLC) protocol.  This protocol provides a packet header for stream synchronization and a packet checksum to validate correct reception, but the HDLC protocol does not provide forward error correction.  Any single bit error will cause the checksum to fail and be counted as a packet loss.   In practice, AIS re-transmits every few seconds depending on the speed and size of the vessel and is tolerant to lost packets from weak signals at far distances.  Commercial AIS systems operate with ranges approaching 100~km.

The synthetic AIS packets used in this experiment were generated from a collection of recorded NMEA payloads obtained along the South Carolina coast. These packets were then regenerated at a high repetition rate to facilitate rapid measurement of the packet loss rate. The probability of detection is estimated by calculating the percentage of dropped packets within a specific time interval at a given transmission rate and electric field strength. The stimulated electric field strength from these packets is controlled by scaling the floating-point amplitude of the GMSK modulation according to the calibration scale. For example, the calibration shows that for a 20~dB gain on the radio, we see a field of roughly 7~V/m on the atoms.

\section{Results and Discussion}

\begin{figure}[!ht]
    \centering
    \includegraphics[width=.48\textwidth]{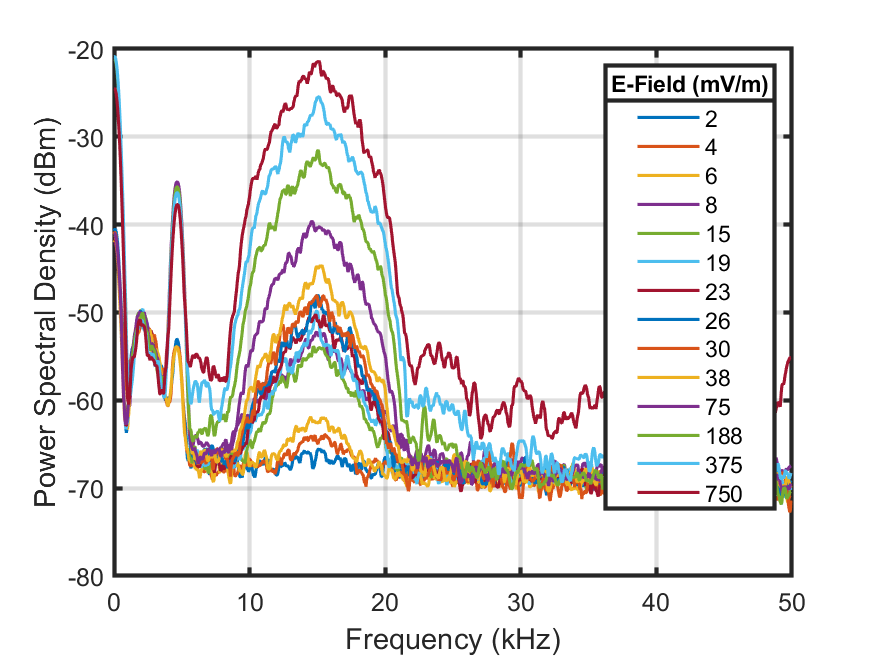}
    \caption{Sample spectrum of AIS signal reception using atoms. The AIS signal is 10 kHz from the LO with a 9600 Hz Gaussian minimal shift keying modulation. Different traces correspond to different levels of received field strength.}
    \label{fig:sample_spectrum}
\end{figure}

The comparison of the AC Stark shifting and the HAMMER method are done by looking at two parameters: sensitivity (receiver noise floor) and 10\% packet success rate of each detection method.
The sensitivity is measured using a spectrum analyzer set to 470 Hz resolution bandwidth. 
We obtain a spectrum of the AIS Gaussian packet spread across the 9600~Hz bandwidth that is received by the atoms, an example of which as measured with the Rb atoms is shown in Fig.~\ref{fig:sample_spectrum}.  We define the sensitivity as the incident electric field strength that causes the peak received power spectral density to be equal that of the receiver noise floor. In our system, the noise floor is set by a low frequency thermal atomic noise that sits 10~dB above the shot noise of the laser, similar to what is seen in \cite{Repump_paper}.

\begin{figure}[!ht]
    \centering
    \includegraphics[width=.48\textwidth]{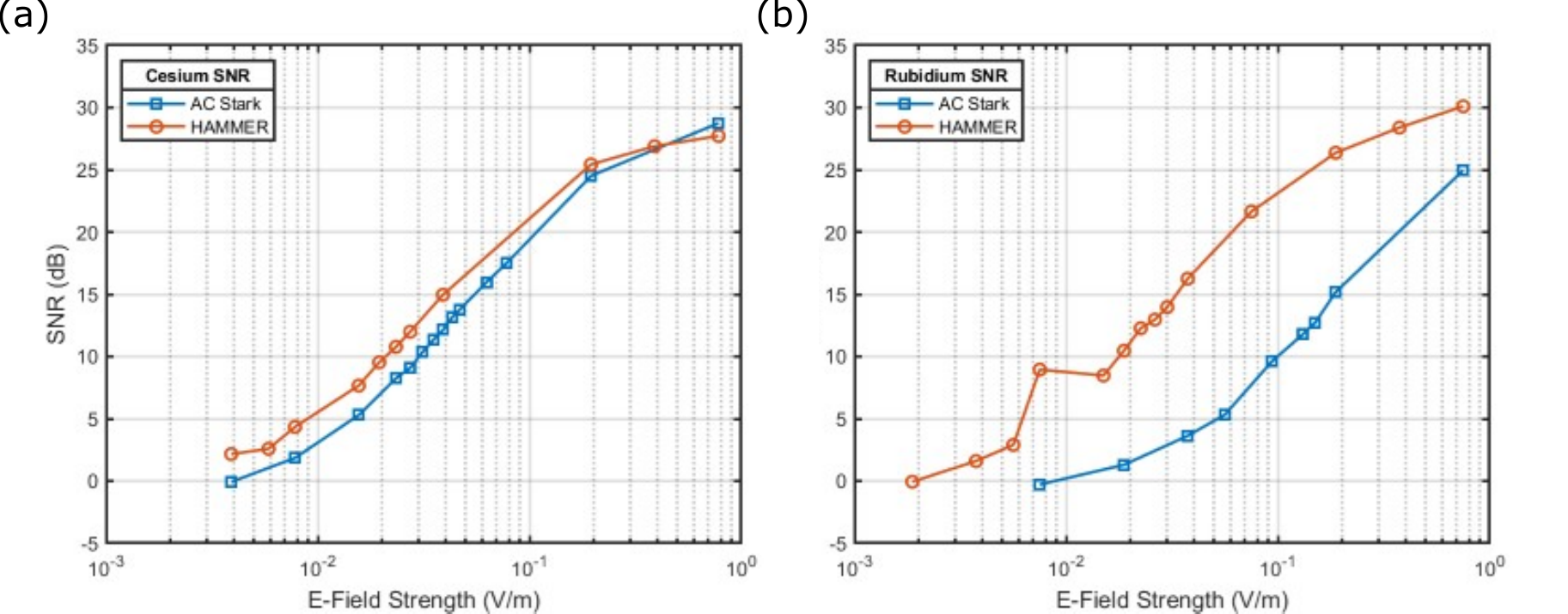}
    \caption{signal-to-noise (SNR) of field detected to detected noise (dominated by laser and thermal atomic noise) as a function of calibrated field strength. (a) Data for Cs atoms for the two methods, AC Stark shifting (blue) and HAMMER method (red). (b) Same as (a), but for Rb atoms.}
    \label{fig:sensitivity}
\end{figure}

The signal-to-noise ratio (SNR) was obtained by taking the ratio of signal and noise power spectral densities.  That is, the ratio of the total received power integrated across the spectral mask and divided by the bandwidth of the spectral mask (to estimate the received signal power spectral density) to that of the noise power spectral density in the region adjacent to the spectral mask. The SNR was recorded at various electric field stimulation levels. Fig.~\ref{fig:sensitivity} depicts the relationship between stimulated electric field strength and the observed SNR on the spectrum analyzer for the different atomic species and methods used.
Notably, in both Rb and Cs, the HAMMER detection method demonstrates superior sensitivity compared to AC Stark shifting detection for the 162~MHz signal.  

The improved sensitivity is caused by the large polarizability of the Rydberg G state. For the case of Rb, there is a large discrepancy between the polarizability of the 50D$_{5/2}$ state and the 48G$_{9/2}$ state. The polarizability for 50D$_{5/2}$ is 0.02~$\mathrm{GHz(V/m)^{-2}}$ while the polarizability for the 48G$_{9/2}$ state is roughly 20~$\mathrm{GHz(V/m)^{-2}}$. This large polarizability of the 48G$_{9/2}$ state elicits a strong response from the atoms. However, because the G state is reached through the coupling of two RF resonances, line misshaping and broadening from the Stark tuning of the G state resonance overall reduces the full benefits to sensitivity that would be possible with this method. For the case of Cs, there is a less pronounced difference in the polarizability. This is due to the already large polarizability of the Rydberg 55D$_{5/2}$ state that the lasers excite to. The polarizability of the Cs 56D$_{5/2}$ state is 0.4~$\mathrm{MHz(V/m)^{-2}}$ while the polarizability of the Cs 54G$_{9/2}$ state is roughly 10~$\mathrm{MHz(V/m)^{-2}}$. Furthermore, the separation of the F and G states in Rb is 700~MHz while it is nearly 1100~MHz in Cs. Because the Cs atoms require a larger Stark shift to move the G state and F state to be resonant with the 162~MHz AIS field, there can be more inhomogeneous broadening effects from nonuniform fields and state mixing. These effects also effect the Rb system, but since there is a larger polarizability and smaller separation of the F and G states, this effect is reduced.

The best sensitivity that we were able to measure was from the Rb system using the Hammer method. Fig.~\ref{fig:sensitivity} (b) shows that the 0~dB SNR location occurs with 2~mV/m signal field. To estimate the sensitivity for the system, we integrate the power spectral density that is spread across the 9600 Hz of bandwidth to find the single tone equivalent power and field. For simple signals like QPSK, this would simply be a linear change in the power with respect to bandwidth. So roughly 40 dB change for 9600 Hz. However, for the Gaussian profile, this integration results in a 26 dB change to the SNR. For a representative single tone, our sensitivity is then 100~$\mu$V/m/$\sqrt{Hz}$. However, in recent tests with different beam sizes and optical powers, we have managed to observe this level of sensitivity with just the Stark shifting method in cesium. This leads us to believe the Hammer could potentially lead to sensitivity on the order of 30~$\mu$V/m/$\sqrt{Hz}$.

While we may try to make a comparison between Cs and Rb, this is simply to compare the benefits of the HAMMER detection method rather than overall sensitivity of the two systems. On inspection, it appears that Rb exhibits superior sensitivity compared to Cs, shown by Fig.~\ref{fig:sensitivity}. However, it is important to consider that the vapor cell dimensions, volumes, and pressures differ between Cs and Rb. Therefore, these experiments conducted with Cs and Rb should not be directly compared to evaluate the relative sensitivity of each species.

\begin{figure}[!ht]
    \centering
    \includegraphics[width=.48\textwidth]{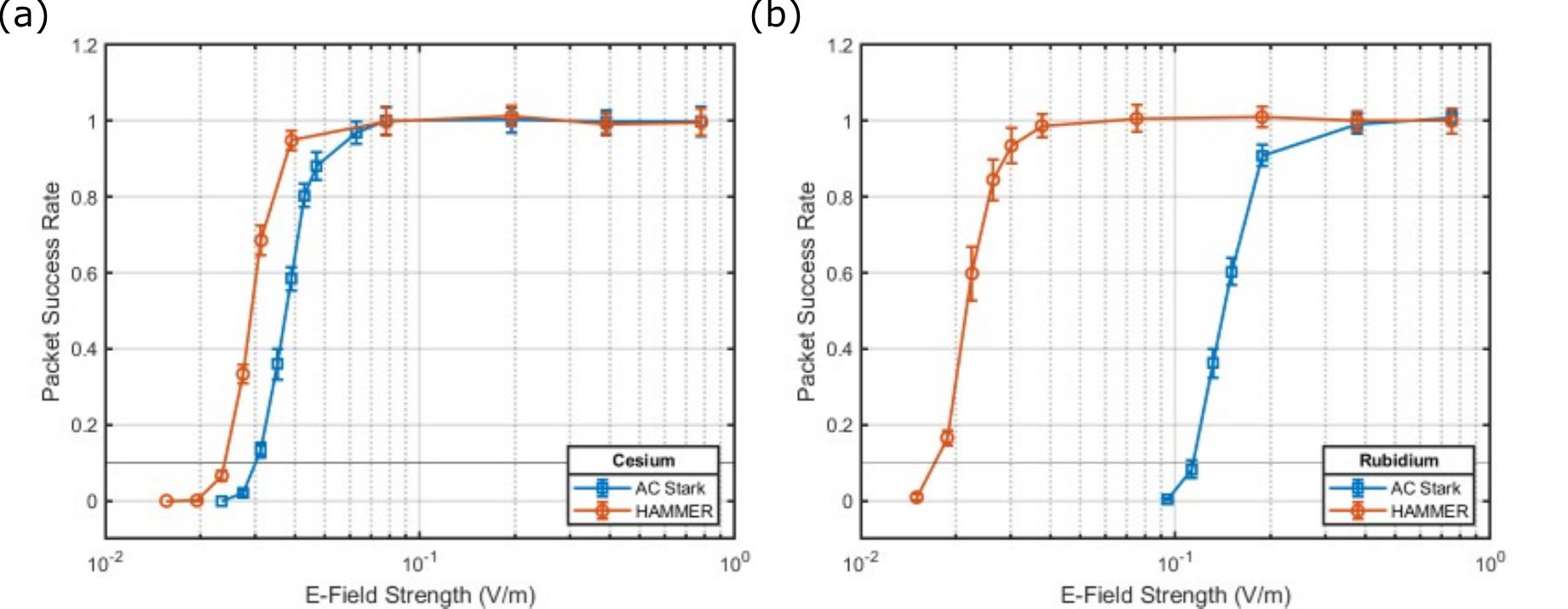}
    \caption{Packet success rate as a function of field strength. (a) Data for Cs atoms for the two methods, AC Stark shifting (blue) and HAMMER method (red). (b) Same as (a), but for Rb atoms.}
    \label{fig:packet_success}
\end{figure}
The probability of successfully decoding an AIS packet is determined by the SNR at the receiver. Fig.~\ref{fig:packet_success} illustrates the observed rate of successful packet detections in the software-defined radio as a function of the incident electric field strength. Using a 10\% probability of detection as the operational threshold, a field strength of 17~mV/m is required in Rb using HAMMER and 114~mV/m is required using AC Stark shifting. In Cs, a field strength of 24~mV/m is needed using HAMMER, while 30~mV/m is required using AC Stark. 

At the 10\% packet detection threshold for electric field strength, the SNR based on the spectrum analyzer observations is 11~dB. Appendix~\ref{sec:AppendSNR} provides an analysis that supports the reasonableness of this result given the radio hardware and AIS modulation protocol. For comparison, an 11~dB SNR in an optimized commercial AIS receiver with a noise figure of 4~dB would correspond to an incident electric field strength of approximately 1~$\mu$V/m. On the other hand, the observed minimal detectable field for 10\% success in Rb of 17~mV/m with 11 dB of SNR is an equivalent RF receiver with a noise figure of 86~dB.

\begin{figure}
    \centering
    \includegraphics[width=.48\textwidth]{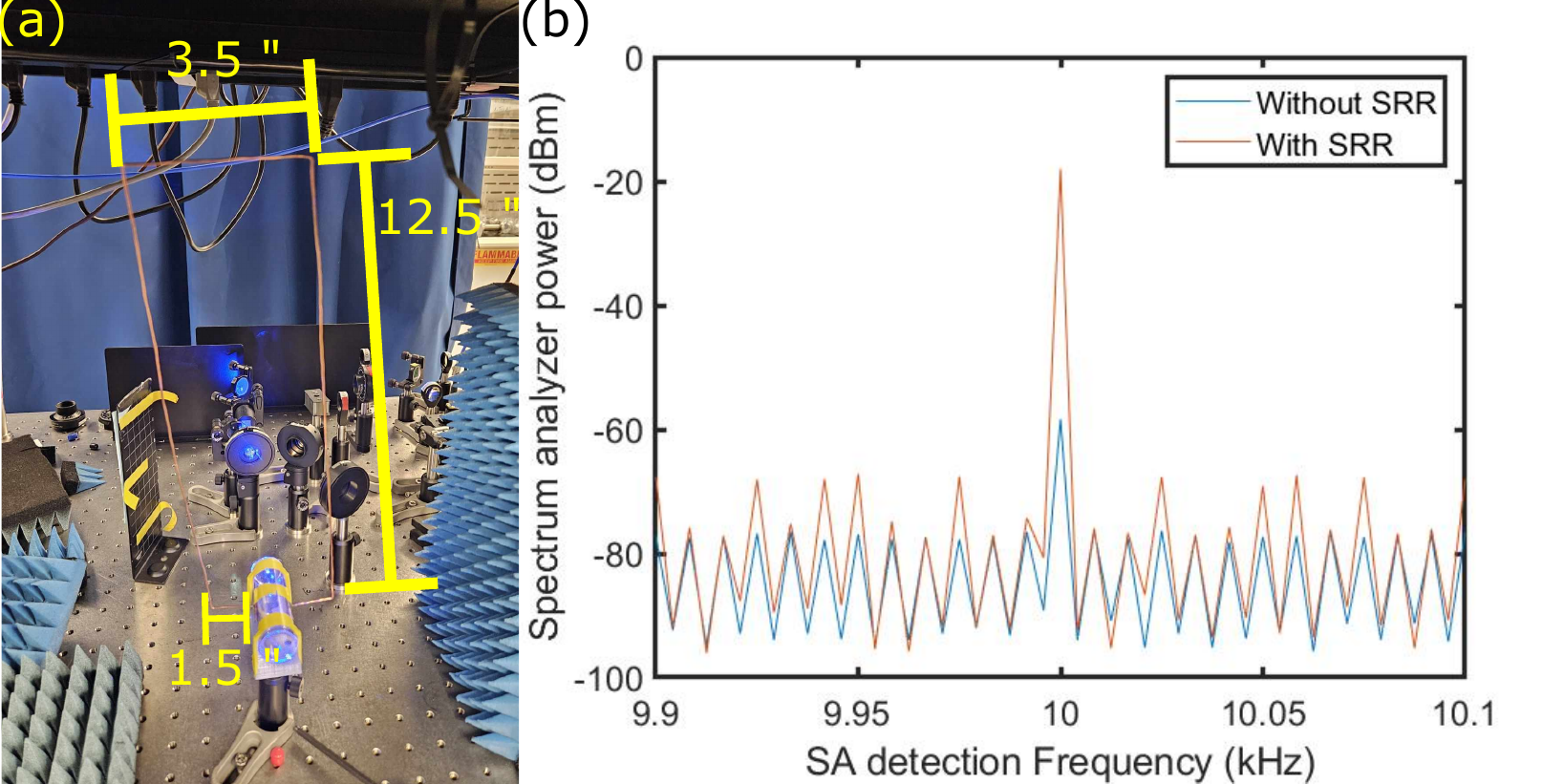}
    \caption{(a) Split-ring resonator structure dimensions and cell. (b) Single tone beat note between an LO and a signal received by the spectrum analyzer. (red) Trace showing data with the SRR present. (blue) Trace showing data without the SRR present.}
    \label{fig:SRR}
\end{figure}

Despite the relatively poor sensitivity at 162 MHz, we suspect that adding resonant structures to the vapor cell may enhance the electric field sensitivity within a narrow band, similar to an optimized classical receiver antenna. These types of structures have been successfully utilized in previous studies~\cite{SRR}, and they hold the potential to improve signal performance. We demonstrate this improvement by making a split-ring resonator (SRR) centered at 162 MHz.shown in Fig.~\ref{fig:SRR} (a). We were not able to obtain a calibrated measurement due to the antenna power limitations, but managed to show a factor of 40~dB enhancement in signal when we compared the signal strength received for with and without the SRR, shown in Fig.~\ref{fig:SRR} (b). This result suggest a SRR will make the sensor competitive with conventional receivers.
\begin{figure}
    \centering
    \includegraphics[width=.48\textwidth]{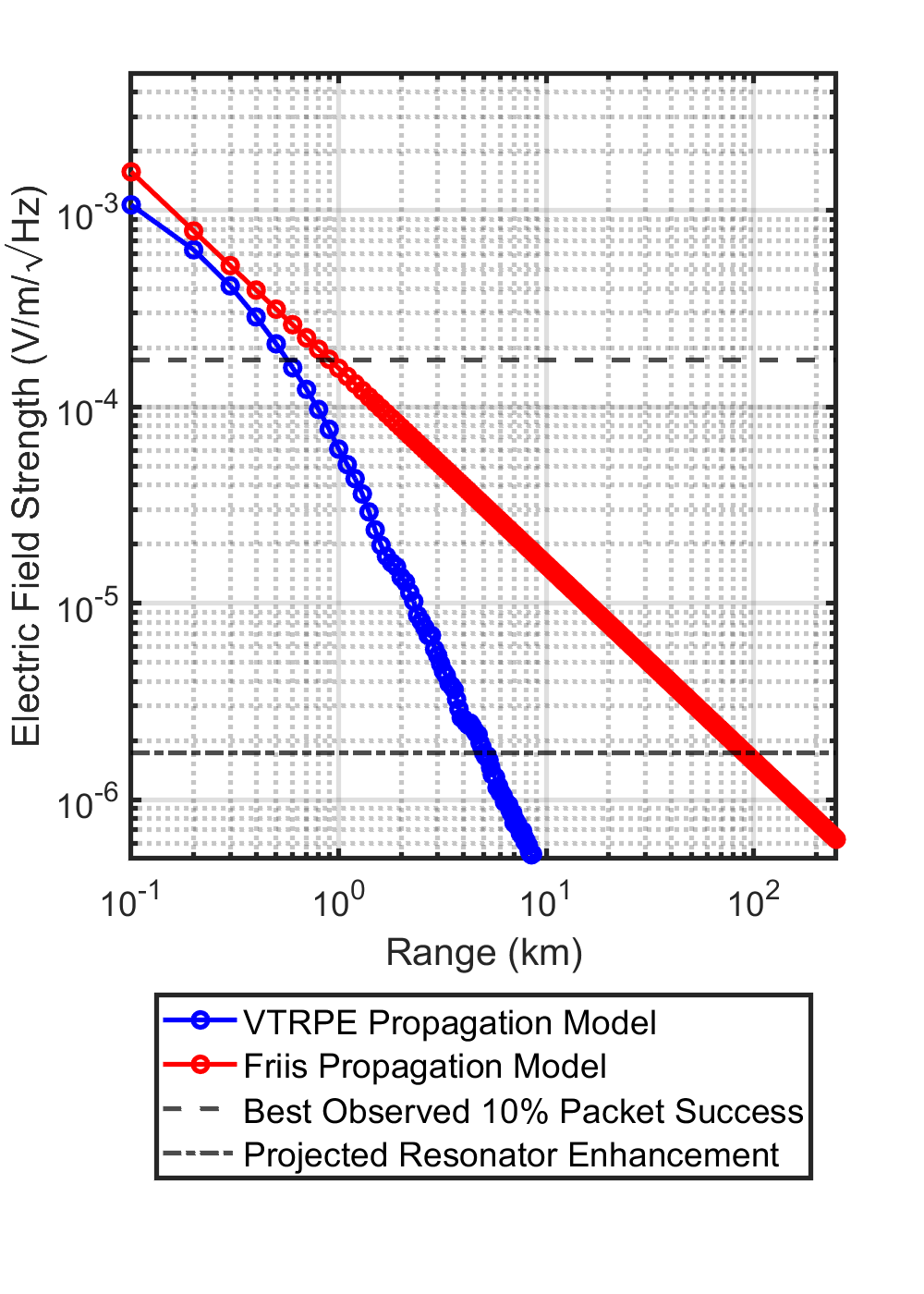}
    \caption{Modeled electric field strength obtained from a Class A AIS transmitter at 5~meters elevation with a quarter wave monopole antenna radiating over water.}
    \label{fig:range_est}
\end{figure}

For the end user, the most critical performance criterion of the AIS system is its effective range. This can be predicted by utilizing propagation models and the transmitter specifications of the AIS system to estimate the range corresponding to a given electric field strength. In this study, we present two models for range prediction. The Variable Terrain Radio Parabolic Equation (VTRPE) model~\cite{Devi2013BERPO} is a numerical method developed by Naval Information Warfare Center Pacific that models RF propagation over physical terrain maps. The Friis transmission model\cite{balanis2012antenna}, on the other hand, is a simpler model that assumes equally dispersed radiation over a sphere. For an electric field strength of 17~mV/m, the minimum detectable field in the Rb HAMMER system that returned at least 10\% packet success rate, the predicted effective range approaches 1~km, depending on the propagation model used. Fig.~\ref{fig:range_est} provides a comparison between the higher fidelity VTRPE model and the standard Friis model for a Class A 12.5~W transmitter spread over 9600~Hz operating in open sea conditions at 5~meters elevation with a quarter wave monopole antenna.

\section{Conclusion}
In this study, we explored the application of Rydberg atom electrometry for radio-frequency communications, particularly in the context of automated identification system (AIS) waveforms used in maritime navigation. Through experiments with cesium and rubidium vapor cells, we observed the potential of utilizing high-angular Momentum Matching Excited Raman (HAMMER) detection for enhanced low-frequency detection and superior sensitivity compared to traditional AC Stark effect detection. The results demonstrated the relationship between incident electric field strength and signal-to-noise ratio (SNR), providing insights into the performance of the atomic vapor cell antenna. Moreover, we assessed the range prediction using propagation models and concluded the current technology provides less than 1~km of operational range. Future research could explore the integration of resonant structures into the vapor cell to further enhance electric field sensitivity. Overall, these findings contribute to the understanding of Rydberg atom electrometry and its potential applications in VHF radio frequency communications.
\appendix

\section{Field Calibration Measurements}\label{sec:AppendCal}

The Rb and Cs calibration was done with the same conditions for the plates and other experimental parameters using the Rydberg D$_{5/2}$ state (with different n). Even with the very large differences in possibilities between the Cs system and the Rb system, their calibration agrees to within a factor of 2. The Rb shifts were difficult to observe and the rough calibration can be attributed to the ratio of the shift expected and the linewidth of the two-photon EIT resonance.

\begin{figure}[!ht]
    \centering
    \includegraphics[width=.48\textwidth]{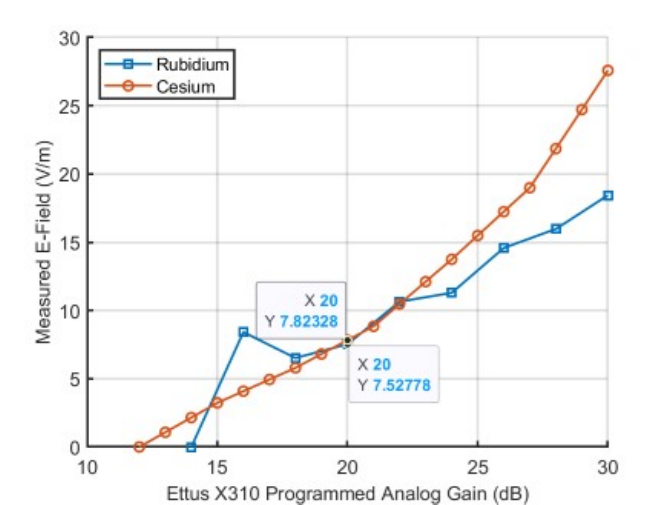}
    \caption{Electric field strength obtained from AC Stark shift produced by the Ettus X310 SDR when driving parallel copper plates around the vapor cell under a constant tone of 0.5 floating point amplitude.  The 20 dB gain was used in all the experiments while varying the floating-point amplitude.}
    \label{fig:calibration}
\end{figure}

\section{SNR Packet Loss Basis}\label{sec:AppendSNR}
Some readers may find it unusual to have an SNR of 11~dB with a 10\% probability of packet detection. However, a quick theoretical calculation justifies this value. An AIS packet typically contains a maximum of 168 bits, including protocol overhead and preamble symbols. Without error coding, the probability of successful detection can be modeled as a series of 168 independent Bernoulli trials. Assuming a success rate of 0.10, we find that the corresponding bit error probability is approximately 0.0136.
According to Ref.~\cite{Devi2013BERPO}, an SNR of about 8.6~dB inside the software defined radio (SDR) corresponds to this bit error rate. The quantization noise of the Ettus X310 radio, with a UBX-160 and an effective number of bits of 11.3, dominates the noise floor at -95.81 dBm/Hz within the radio. The homodyne receiver in this radio has an low-noise amplifier (LNA) with 28~dB of gain, a mixer with a conversion loss of 5.2~dB, and bandpass filters with a net loss of 3~dB. The radio's internal analog gain of 17~dB results in a net receiver gain of 36.8~dB. Subtracting this gain from the ADC noise floor gives an input-referred noise floor of -132.6~dBm/Hz. This is approximately 2.4~dB higher than the noise level observed on the spectrum analyzer, which is -135~dBm/Hz.
The 2.4~dB difference between the ADC quantization noise and the noise floor on the spectrum analyzer accounts for the predicted 8.6~dB SNR and the recorded 11~dB SNR. In simple terms, the radio's analog-to-digital conversion introduces additional noise that is not reflected in the spectrum analyzer plots, thus justifying the observations.

\newpage
\section*{Acknowledgements}
\vspace{-5mm}
\noindent This work was partially funded by the NIST-on-a-Chip (NOAC) Program and was developed with funding from the the Naval Information Warfare Center's NISE program. 

\vspace{-5mm}
\subsection*{Conflict of Interest}
\vspace{-5mm}\noindent The authors have no conflicts of interests to disclose.

\vspace{-5mm}
\subsection*{Data Availability Statement}
\vspace{-5mm}\noindent The data relevant to the findings of this research project are available from the corresponding author upon reasonable request.

\end{document}